\documentclass[12pt,twoside]{article}
\usepackage{a4wide,latexsym,graphicx,epsfig,psfrag}

\usepackage{graphics}
\usepackage{epsfig}

{\catcode `\@=11 \global\let\AddToReset=\@addtoreset}
\AddToReset{equation}{section}
\renewcommand{\theequation}{\thesection.\arabic{equation}}

\def\greaterthansquiggle{\raise.3ex\hbox{$>$\kern-.75em\lower1ex\hbox{$\sim$}}}
\def\lessthansquiggle{\raise.3ex\hbox{$<$\kern-.75em\lower1ex\hbox{$\sim$}}}
\newcommand{\beq}{\begin{equation}}
\newcommand{\eeq}{\end{equation}}
\newcommand{\beqa}{\begin{eqnarray}}
\newcommand{\eeqa}{\end{eqnarray}}
\newcommand{\beqan}{\begin{eqnarray*}}
\newcommand{\eeqan}{\end{eqnarray*}}
\newcommand{\ba}{\begin{array}}
\newcommand{\ea}{\end{array}}
\newcommand{\no}{\nonumber}

\newcommand{\ra}{\rightarrow}

\newcommand{\ve}{\varepsilon}

\newcommand{\bea}{\begin{eqnarray}}
\newcommand{\eea}{\end{eqnarray}}

\def\nz{\ifmmode {I\hskip -3pt N} \else {\hbox {$I\hskip -3pt N$}}\fi}
\def\zz{\ifmmode {Z\hskip -4.8pt Z} \else
       {\hbox {$Z\hskip -4.8pt Z$}}\fi}
\def\qz{\ifmmode {Q\hskip -5.0pt\vrule height6.0pt depth 0pt
       \hskip 6pt} \else {\hbox
       {$Q\hskip -5.0pt\vrule height6.0pt depth 0pt\hskip 6pt$}}\fi}
\def\rz{\ifmmode {I\hskip -3pt R} \else {\hbox {$I\hskip -3pt R$}}\fi}
\def\cz{\ifmmode {C\hskip -4.8pt\vrule height5.8pt\hskip 6.3pt} \else
       {\hbox {$C\hskip -4.8pt\vrule height5.8pt\hskip 6.3pt$}}\fi}

\def\au{{\setbox0=\hbox{\lower1.36775ex%
\hbox{''}\kern-.05em}\dp0=.36775ex\hskip0pt\box0}}
\def\ao{{}\kern-.10em\hbox{``}}

\normalsize
\begin{document}
\bibliographystyle{plain}
\begin{titlepage}
\begin{flushright}
LA-UR 11-00400\\
UWThPh-2011-4 \\
January 2011
\end{flushright}
\vspace{2.5cm}
\begin{center}
{\Large \bf A note on isospin violation in  \boldmath $ {P_{\ell 2 
(\gamma)}}$ \unboldmath
decays
}
\\[40pt]
Vincenzo Cirigliano$^{1}$, Helmut Neufeld$^{2}$

\vspace{1cm}
${}^{1)}$ Theoretical Division, Los Alamos National Laboratory,
Los Alamos, NM 87545, USA \\[10pt]

${}^{2)}$ University of Vienna, Faculty of Physics, 
Boltzmanngasse 5, A-1090 Wien, Austria

\vfill
{\bf Abstract} \\
\end{center}
\noindent
We discuss the size of isospin violating effects of 
both electromagnetic ($e \ne 0$) and strong ($m_u \ne m_d$) 
origin in $P_{\ell 2 (\gamma)}$ decays. Our results are relevant for 
the extraction of $|V_{us}/V_{ud}|$ from the measured ratio of the  
$K_{\mu 2 (\gamma)}$ and $\pi_{\mu 2 (\gamma)}$ decay widths
combined with $F_K/F_\pi$ obtained in lattice 
calculations in the isospin limit. 
We point out that strong isospin-breaking corrections,   
neglected in previous studies, 
enter at the same level as the electromagnetic effects. 
The updated value for the ratio of CKM elements reads 
$|V_{us}/V_{ud}| =  0.2316(12)$.

\vfill
%

\end{titlepage}
\section{Introduction}
\label{sec: Introduction}
\renewcommand{\theequation}{\arabic{section}.\arabic{equation}}
\setcounter{equation}{0}

The purely leptonic decays of light pseudoscalar mesons $P \to \ell  \nu_\ell$ 
($P^\pm = \pi^\pm, K^\pm$) are theoretically extremely clean.  
Neglecting electroweak corrections,  the hadronic physics is encoded in 
one single parameter, the meson decay constant $F_P$. 
Using  chiral perturbation theory and lattice QCD,  $P \to \ell \nu_\ell$ 
decay rates  can be predicted with high accuracy and provide non-trivial tests 
of the standard model imposing stringent constraints on new physics.  

With the advent of high precision calculations of the ratio of decay constants 
$F_K/F_\pi$ in lattice QCD, it was realized  \cite{Marciano:2004uf} that 
the combination of $F_K/F_\pi$ and the ratio of measured 
decay widths 
$\Gamma_{K_{\mu 2 (\gamma)}} / \Gamma_{\pi_{\mu 2 (\gamma)}}$ 
leads to a clean and competitive  determination of  the ratio 
$|V_{us}/V_{ud}|$ of 
Cabibbo-Kobayashi-Maskawa (CKM)~\cite{Cabibbo:1963yz,Kobayashi:1973fv} 
matrix elements.
Implementing the programme proposed in Ref.~\cite{Marciano:2004uf} 
requires theoretical understanding of:   
(i) the long-distance  electromagnetic (EM)  corrections 
to $P \to \ell  \nu_\ell$;  
(ii)  the  strong isospin breaking  correction that  connects $F_K/F_\pi$ 
(calculated within lattice QCD  in the isospin limit $m_u=m_d$)
to  $F_{K^\pm}/F_{\pi^\pm}$ that enters the ratio of 
$K_{\mu 2 (\gamma)}$ to  $\pi_{\mu 2 (\gamma)}$  decay widths. 

The EM  corrections  have been known for some time  
\cite{Marciano:1993sh,Sirlin:1977sv,Sirlin:1981ie,Finkemeier:1995gi}.
The analysis of Ref.~\cite{Marciano:2004uf}  relied on the model-dependent 
treatment of Ref.~\cite{Finkemeier:1995gi}  
for the long-distance contribution to these corrections. 
The long-distance EM corrections to $P \to \ell \nu_\ell (\gamma)$ 
have also been analyzed  in the model-independent framework of 
Chiral Perturbation Theory 
(ChPT)~\cite{Knecht:1999ag,Cirigliano:2007ga,Gasser:2010wz}. 
It was noted in Ref.~\cite{Knecht:1999ag} that the radiative corrections 
to the ratio
$\Gamma_{K_{\mu 2 (\gamma)}} / \Gamma_{\pi_{\mu 2 (\gamma)}}$ 
are uniquely determined to lowest non-trivial order in the chiral 
expansion and the final numerical results have been reported in Refs.
\cite{Antonelli:2009ws,Antonelli:2010yf}. 
So far, strong isospin breaking corrections $\propto (m_u -m_d)$  have 
been ignored in the extraction of $|V_{us}/V_{ud}|$. In view of the 
increasing precision of lattice results on $F_K/F_\pi$, also these effects 
should be taken into account and 
it is the purpose of this letter to fill this gap.

In Section~\ref{sec: decays}
we perform our analysis  of both electromagnetic  and strong isospin 
violation to 
 $\Gamma_{P_{\ell 2 (\gamma)}}$   in the unified framework of ChPT. 
In Section~\ref{sec:pheno} we update  the phenomenological determination 
of $|V_{us}/V_{ud}|$ 
and the global fit to $|V_{ud}|$ and $|V_{us}|$~\cite{Antonelli:2010yf}
to reflect the new strong isospin breaking correction.
We present our conclusions in Section~\ref{sec: Conclusions}.

\section{Isospin violation in 
\boldmath 
$\pi \ra \ell \nu_{\ell} (\gamma)$ 
\unboldmath
and 
\boldmath 
$K \ra \ell \nu_{\ell} (\gamma)$
\unboldmath}
\label{sec: decays}
\renewcommand{\theequation}{\arabic{section}.\arabic{equation}}
\setcounter{equation}{0}

The fully photon-inclusive decay width
\begin{equation}
\Gamma_{P_{\ell 2 (\gamma)}} \equiv
\Gamma(P \ra \ell \nu_{\ell}) 
+ \Gamma(P \ra \ell \nu_{\ell} \gamma)~,
\quad P = \pi^\pm,  K^\pm~,
\end{equation}
can be written in the form \cite{Marciano:1993sh,Cirigliano:2007ga}
\begin{equation} \label{Pl2g}
\Gamma_{P_{\ell 2 (\gamma)}} 
=  
\frac{G_F^2 |V_P|^2  F_P^2 }{4 
\pi} \, 
M_P  \, m_\ell^2  \, \left(1 - \frac{m_\ell^2}{M_P^2} \right)^{\! 2}
\! S_{\mathrm {EW}} \, \left( 1 + \delta_{\mathrm {EM}}^{P} \right)
~,
\quad
V_{\pi^\pm} = V_{ud}~, \, V_{K^\pm} = V_{us}~. 
\end{equation}
$G_F$ is the Fermi constant extracted from muon decay and $F_{\pi^\pm}$ 
($F_{K^\pm}$) is the decay constant of the charged pion (kaon) in pure 
QCD\footnote{In the case of the pion, the distinction between the charged 
and the neutral decay constant is a tiny effect of order $(m_d-m_u)^2$, 
whereas $F_{K^\pm}/F_{K^0}$ is of the order $m_d-m_u$ \cite{Gasser:1984gg}. 
For a discussion of some subtle points in the separation of QCD and electromagnetism, the 
reader is referred to Refs. \cite{Knecht:1999ag}, \cite{Gasser:2003hk}, 
and \cite{Gasser:2010wz}. 
Here we simply recall that the scale ambiguity inherent to the separation 
of QCD and QED effects in the meson decay constants in the chiral limit 
has been shown to be about $8$ keV~\cite{Gasser:2010wz}, and thus negligible 
at the current level of precision.}.
 $S_{\mathrm 
{EW}} = 1.0232$ is the universal electroweak short-distance enhancement 
factor \cite{Marciano:1993sh,Sirlin:1977sv,Sirlin:1981ie}  appearing 
in semileptonic decays. 
The long-distance electromagnetic corrections are described by 
\cite{Marciano:1993sh}
\begin{equation}
\delta_{\mathrm {EM}}^P = 
\frac{\alpha}{\pi}  \left(   F (m_\ell/M_P) 
+ \frac{3}{4}  \ln \frac{M_P^2}{M_\rho^2}  
- C_1^P
+ \ldots
\right) ~.  
\end{equation}
The one-loop function $F(x)$ is related to the universal long-distance
electromagnetic corrections (limit of point-like meson), its explicit
form can be found in Ref. \cite{Marciano:1993sh}. The dimensionless 
constant
$C_1^P$ parametrizes electromagnetic effects due to hadronic 
structure. The dots refer to further structure dependent terms 
\cite{Cirigliano:2007ga} arising at higher orders in the chiral expansion.

The constants
$C_1^\pi$ and $C_1^K$ were calculated 
in chiral perturbation theory with virtual photons and leptons
to order $e^2 p^2$ \cite{Knecht:1999ag}:
\begin{eqnarray}
C_1^{\pi}
&=&
  - \tilde{E}^r(M_\rho) + 
\frac{Z}{4} \left( 3 +2 \ln \frac{M_\pi^2}{M_\rho^2} + \ln 
\frac{M_K^2}{M_\rho^2} \right) ~,
\\
C_1^{K}
&=&
  -  \tilde{E}^r(M_\rho)  + 
\frac{Z}{4} \left( 3 +2 \ln \frac{M_K^2}{M_\rho^2} + \ln 
\frac{M_\pi^2}{M_\rho^2} \right) ~.
\end{eqnarray}
The parameter $Z$, appearing in the chiral Lagrangian of order $e^2 p^0$, can 
be expressed in terms of the measured pion masses, 
the fine-structure constant and the pion decay constant: 
\begin{equation}
Z = \frac{M_{\pi^\pm}^2 - M_{\pi^0}^2}{8 \pi \alpha F_\pi^2} \simeq 0.8~.
\end{equation} 
Both $C_1^\pi$ and  $C_1^K$ depend on the same combination 
of electromagnetic low-energy couplings,
\begin{equation} \label{Eschlange}
\tilde{E}^r
=\frac{1}{2} + 4 \pi^2 \left(
\frac{8}{3} K_1^r 
+ \frac{8}{3} K_2^r
+ \frac{20}{9} K_5^r 
+ \frac{20}{9} K_6^r 
-\frac{4}{3} X_1^r
- 4 X_2^r
+ 4 X_3^r
- \tilde{X}_6^{\rm phys} \right)~.
\end{equation}
The $K_i$ were defined in Ref. \cite{Urech:1994hd} and the $X_i$ in 
Ref. \cite{Knecht:1999ag}. 
Note that we have pulled out 
the short-distance enhancement factor in Eq. (\ref{Pl2g}) 
keeping only the residual long-distance part
$\tilde{X}_6^{\mathrm {phys}}$ 
\cite{DescotesGenon:2005pw} 
in Eq. (\ref{Eschlange}). The coupling constants entering in 
Eq. (\ref{Eschlange}) 
have been estimated in Refs. \cite{Ananthanarayan:2004qk} and 
\cite{DescotesGenon:2005pw} using large-$N_c$ methods, giving
\begin{equation}
C_1^\pi = -2.4(5)~, \quad
C_1^K = -1.9(5)~.
\end{equation}
The errors given here are based on naive power counting of unknown 
contributions arising at order $e^2 p^4$. 

For the ratio of decay widths we find 
\begin{equation} \label{wratio}
\frac{\Gamma_{K_{\ell 2 (\gamma)}}}{\Gamma_{\pi_{\ell 2 (\gamma)}}}
=
\frac{|V_{us}|^2}{|V_{ud}|^2} \,
\frac{F_{K^\pm}^2}{F_{\pi^\pm}^2} \, 
\frac{M_{K^\pm} \left( 1-m_\ell^2/M_{K^\pm}^2 \right)^2}
{M_{\pi^\pm} \left( 1-m_\ell^2/M_{\pi^\pm}^2 \right)^2} \,
\left( 1+\delta_{\mathrm {EM}} \right)~,
\end{equation}
where
\begin{eqnarray} \label{deltaEM}
\delta_{\mathrm {EM}} &=& 
\delta_{\mathrm {EM}}^K -
\delta_{\mathrm {EM}}^\pi 
\nonumber \\
&=&
\frac{\alpha}{\pi}
\left(
F(m_\ell/M_K) - F(m_\ell/M_\pi)
+ \frac{3}{4} \ln \frac{M_K^2}{M_\pi^2} 
-C_1^K + C_1^\pi
\right)
\nonumber
\\
&=&
\frac{\alpha}{\pi}
\left(
F(m_\ell/M_K) - F(m_\ell/M_\pi)
+ \frac{3-Z}{4} \ln \frac{M_K^2}{M_\pi^2} 
\right)
\nonumber
\\
&=& -0.0069(17)~. 
\end{eqnarray}
The important feature of the above result is that 
the electromagnetic low-energy coupling $\tilde{E}^r$ 
cancels out in the ratio \cite{Knecht:1999ag}. To lowest non-trivial order 
in the chiral expansion,
the electromagnetic correction $\delta_{\mathrm {EM}}$ is uniquely determined in 
terms of the fine-structure constant and masses,  
demonstrating the power of effective field theory methods.
The quoted $25\%$  uncertainty is an estimate of corrections arising to higher order 
in the chiral expansion. 
Note also that in Eq. (\ref{wratio}) 
 the ratio $F_{K^\pm}/F_{\pi^\pm}$ is the full QCD quantity 
including strong isospin breaking due to $m_u \ne m_d$.

With experimental measurements of the $\pi_{\ell 2 (\gamma)}$ and  
$K_{\ell 2 (\gamma)}$ decay rates and the precise knowledge of the 
radiative corrections
given in  Eq. (\ref{deltaEM}), Eq. (\ref{wratio}) can be used to 
obtain the value of the ratio
\begin{equation} \label{Marciano}
\frac{|V_{us}|}{|V_{ud}|} \, \frac{F_{K^\pm}}{F_{\pi^\pm}} =
\left(
\frac{\Gamma_{K_{\ell 2 (\gamma)}} M_{\pi^\pm}}
{\Gamma_{\pi_{\ell 2 (\gamma)}} M_{K^\pm}}
\right)^{\! 1/2}
\frac
{1-m_\ell^2/M_{\pi^\pm}^2}
{1-m_\ell^2/M_{K^\pm}^2}
\,
\left( 1-\delta_{\mathrm {EM}}/2 \right)~.
\end{equation}
It was suggested in Ref. \cite{Marciano:2004uf} to combine this result 
with 
lattice data on the ratio of
the kaon and pion decay constant to determine $|V_{us}/V_{ud}|$. Finally, 
using also
experimental input for $|V_{ud}|$, a value for $|V_{us}|$ can be obtained in 
this
way. 

As lattice 
calculations~\cite{Beane:2006kx,Follana:2007uv,Aoki:2008sm,Allton:2008pn,Bazavov:2009bb,Blossier:2009bx,Durr:2010hr} 
are usually still performed in the limit of 
equal light quark masses ($m_u = m_d$), it is convenient 
to rewrite Eq. (\ref{wratio}) in the form
\begin{equation} \label{modratio}
\frac{\Gamma_{K_{\ell 2 (\gamma)}}}{\Gamma_{\pi_{\ell 2 (\gamma)}}}
=
\frac{|V_{us}|^2}{|V_{ud}|^2} \,
 \frac{F_K^2}{F_\pi^2}  \, 
\frac{M_{K^\pm} \left( 1-m_\ell^2/M_{K^\pm}^2 \right)^2}
{M_{\pi^\pm} \left( 1-m_\ell^2/M_{\pi^\pm}^2 \right)^2} \,
\left( 1+\delta_{\mathrm {EM}} + \delta_{\mathrm {SU(2)}} \right) ~,
\end{equation}
where $F_\pi$ and $F_K$ denote the decay constants in the isospin limit 
and
\begin{equation}
\delta_{\mathrm {SU(2)}} = 
\left( \frac{F_{K^\pm} F_\pi}{F_{\pi^\pm} F_K} \right)^{\! 2}
-1~.
\end{equation}
The isospin symmetry  limit is defined by  $m_u = m_d$,  $e = 0$ 
and an appropriate  choice  of the isospin symmetric meson masses $M_\pi$, $M_K$.
To the order we are working here, it is sufficient to adopt the following convention: 
$M_\pi$ and $M_K$ are related to the measured values of the masses by 
\begin{equation}
M_\pi^2 = M_{\pi^0}^2~, \quad
M_K^2 = \frac{1}{2} \left(M_{K^\pm}^2 + M_{K^0}^2 - M_{\pi^\pm}^2 + 
M_{\pi^0}^2 \right) ~,
\end{equation}
as the leading order isospin violating contributions of strong and electromagnetic origin 
cancel in these expressions (see e.g. Ref~\cite{Knecht:1999ag}). 
For a fully consistent application of the isospin violating corrections calculated in this work, the lattice QCD 
results should be extrapolated to the isospin-limit ``physical"  meson masses $M_K$ and $M_\pi$. 

The chiral expansion of $F_{\pi^\pm}$ and $F_{K^\pm}$ up to the order 
$p^4, \, 
(m_d-m_u) p^2$ is given by \cite{Neufeld:1995mu}
\beqa \label{Fpipm}
F_{\pi^\pm} &=& F_0 \Bigg\{1 + \frac{4}{{F_0}^2} \left[L_4^r(\mu) 
(M^2_\pi + 2M_K^2)
 + 
L_5^r (\mu) M^2_\pi\right]  \no \\
&& \mbox{} - \frac{1}{2(4\pi)^2 {F_0}^2} \left[ 2 M^2_{\pi} \ln 
\frac{M^2_{\pi}}{\mu^2}  
+ M^2_{K} \ln \frac{M^2_{K}}{\mu^2}\right] \Bigg\}~,
\\
F_{K^{\pm}} &=& F_0 \Bigg\{1 + 
\frac{4}{{F_0}^2} \left[L_4^r(\mu) (M^2_\pi + 2M_K^2)
 + 
L_5^r (\mu) M^2_K\right]  \no \\
&& \mbox{} - \frac{1}{8(4\pi)^2 {F_0}^2} \left[ 3 M^2_{\pi} \ln 
\frac{M^2_{\pi}}{\mu^2}  
+ 6 M^2_{K} \ln \frac{M^2_{K}}{\mu^2} +
3 M^2_{\eta} \ln \frac{M^2_{\eta}}{\mu^2} 
\right] \no \\
&& \mbox{} 
- \frac{8\sqrt{3}\;\varepsilon}{3F_0^2}L_5^r(\mu)(M_K^2-M_\pi^2) 
 \label{FKpm}  \\
&& \mbox{}
-\frac{\sqrt{3}\;\varepsilon}{4(4\pi)^2F_0^2}\left[ 
M^2_{\pi} \ln \frac{M^2_{\pi}}{\mu^2} - 
M^2_{\eta} \ln \frac{M^2_{\eta}}{\mu^2} 
-\frac{2}{3}(M_K^2-M_\pi^2)\left(\ln \frac{M^2_{K}}{\mu^2} + 1\right)
\right]\Bigg\}~.  \no 
\eeqa
$F_0$ is the pion decay constant in the limit of chiral SU(3) and
$M_{\pi}$ and $M_K$ denote the
isospin limits of the pion mass and the
kaon mass, respectively, as discussed above. 
To lowest order, they can be expressed in terms  of 
the quark masses by
\beq
 M^2_{\pi} =  2{B_0} \hat{m}~, \quad 
 M^2_K = {B_0} (m_s + \hat{m} )~, \quad
\hat{m} = \frac{1}{2} (m_u + m_d)~,
\eeq
where $B_0$ is related to the quark condensate.
To the same order, $M_\eta$ is related to $M_K$ and 
$M_\pi$ by 
the GMO relation 
\begin{equation}
M^2_\eta = \frac{4}{3} {B_0}\left( m_s + \frac{\hat{m}}{2}\right)
= \frac{4}{3} M_K^2 - \frac{1}{3} M_\pi^2~.
\end{equation} 
The parameter $\varepsilon$ of strong isospin breaking (which coincides 
with the lowest order $\pi^0$-$\eta$ mixing angle) is given by
\beq
\ve = \frac{\sqrt{3}}{4} \; \frac{m_d - m_u}{m_s - \hat{m}} 
\label{epsilon}~.
\eeq 
Taking $\ve = 0$  in Eqs. (\ref{Fpipm}) and (\ref{FKpm})
reproduces the  results given in Ref. \cite{Gasser:1984gg} in the isospin 
limit. The correction parameter $\delta_{\mathrm {SU(2)}}$  
can be read off as
\begin{eqnarray} \label{deltaSU2v1}
\delta_{\mathrm {SU(2)}} &=&   \mbox{} 
- \frac{16\sqrt{3}\;\varepsilon}{3F_0^2}L_5^r(\mu)(M_K^2-M_\pi^2) 
 \label{FK}  \\
&& \mbox{}
-\frac{\sqrt{3}\;\varepsilon}{2(4\pi)^2F_0^2}\left[ 
M^2_{\pi} \ln \frac{M^2_{\pi}}{\mu^2} - 
M^2_{\eta} \ln \frac{M^2_{\eta}}{\mu^2} 
-\frac{2}{3}(M_K^2-M_\pi^2)\left(\ln \frac{M^2_{K}}{\mu^2} + 1\right)
\right]~. \no
\end{eqnarray}
Expressing $L_5^r$ in terms of the {\it isospin limit} ratio  $F_K/F_\pi$,  $\delta_{\mathrm {SU(2)}}$  
can be recast into the following  compact  form:
\begin{equation} \label{deltaSU2v2}
\delta_{\mathrm {SU(2)}} =    
\sqrt{3} \, \varepsilon
\left[ -\frac{4}{3} \left( F_K / F_\pi -1 \right)
+\frac{1}{3(4\pi)^2F_0^2}
\left( 
M^2_{K} -M_\pi^2 - M_\pi^2 \ln \frac{M^2_K}{M_\pi^2} 
\right) \right]~. 
\end{equation}
With the FLAG \cite{Colangelo:2010et} averages of 
$N_f = 2 + 1$ lattice calculations, 
\begin{equation}
\varepsilon = \frac{\sqrt{3}}{4 R} = 0.0116(13)~, 
\quad
F_K / F_\pi = 1.193(6)~, 
\label{eq:flag}
\end{equation}
we obtain
\begin{equation} \label{deltaSU2num}
\delta_{\mathrm {SU(2)}} =  - 0.0043(5)(11)_{\mathrm {higher \, 
\, orders}}~,
\end{equation}
where we have estimated the uncertainty due to higher order\footnote{
An analysis of isospin violation beyond $O (p^4)$ in 
ChPT~\cite{Amoros:2001cp}  is beyond the scope of this work.}
corrections in the chiral expansion 
to be at a level of $25 \%$.  
We note that the strong isospin-breaking correction
$\delta_{\mathrm {SU(2)}}$   is of the same 
order of magnitude as the electromagnetic 
correction $\delta_{\mathrm {EM}}$   
 in Eq. (\ref{deltaEM}) and should not 
be neglected in the extraction of the ratio  $|V_{us}/V_{ud}|$. 
 
\begin{center}
\begin{figure}[floatfix] 
\centering
\includegraphics[width=15cm]{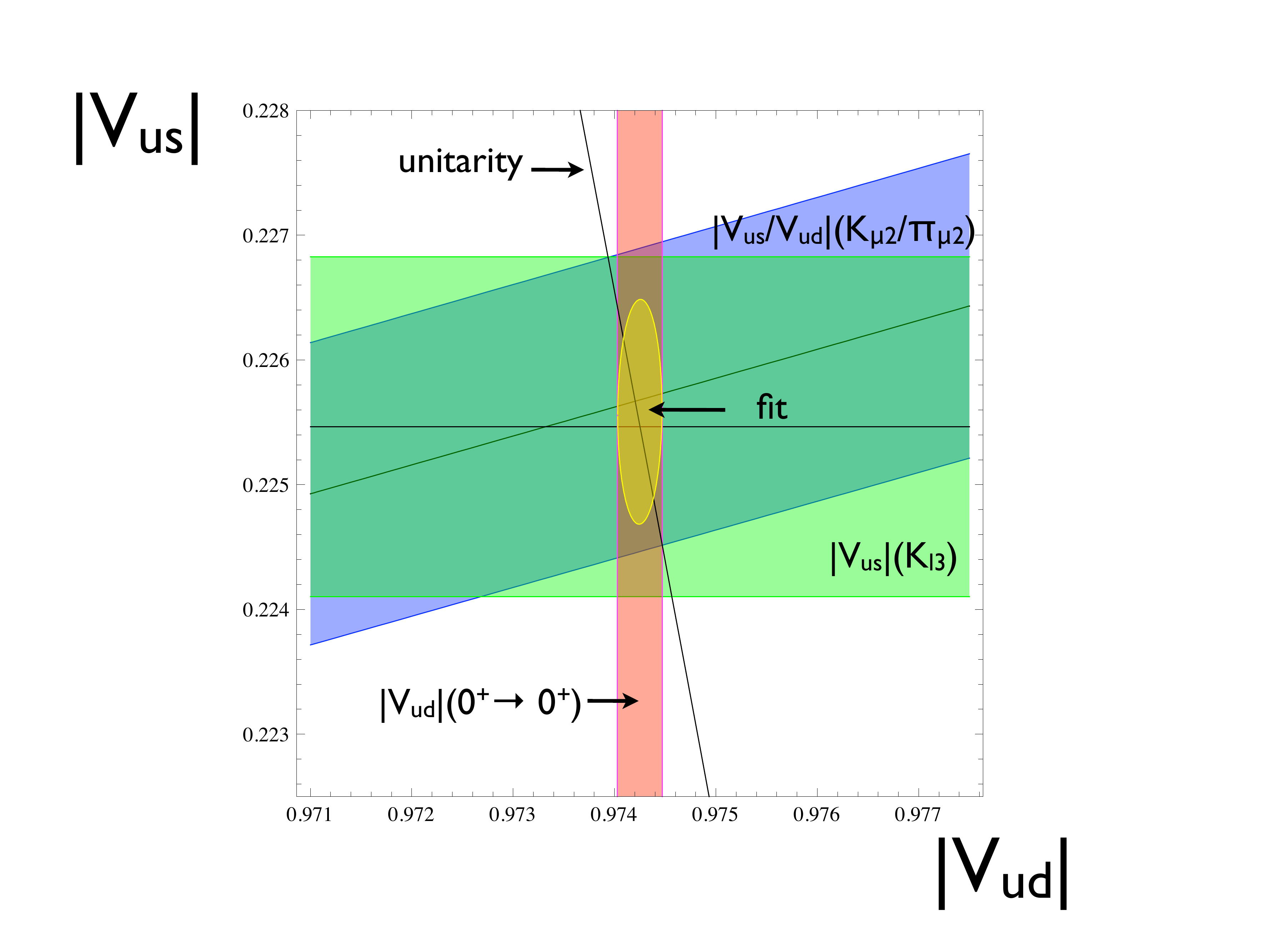}
\caption{\label{fig:unitarity}  Graphical representation of current status 
of $|V_{ud}|$, $|V_{us}|$ and corresponding CKM unitarity test. 
The horizontal band represents the constraint from from $K_{\ell 3}$ decays, 
the thin vertical band the constraint from $0^+ \to 0^+$ nuclear decays, 
the oblique band the constraint from $K_{\mu 2}/\pi_{\mu 2}$, 
and the ellipse is the $1 \sigma$  fit region. 
 }
\end{figure}
\end{center}

\section{Update of \boldmath $|V_{us}/V_{ud}|$ \unboldmath and 
fit to \boldmath $|V_{ud}|, |V_{us}|$ 
\unboldmath }
\label{sec:pheno}
\renewcommand{\theequation}{\arabic{section}.\arabic{equation}}
\setcounter{equation}{0}

We start from the basic formula
\begin{equation} \label{basic-formula}
\frac{|V_{us}|}{|V_{ud}|} \, \frac{F_K}{F_\pi} =
\left(
\frac{\Gamma_{K_{\ell 2 (\gamma)}} M_{\pi^\pm}}
{\Gamma_{\pi_{\ell 2 (\gamma)}} M_{K^\pm}}
\right)^{\! 1/2}
\frac
{1-m_\ell^2/M_{\pi^\pm}^2}
{1-m_\ell^2/M_{K^\pm}^2}
\,
\left( 1-\delta_{\mathrm {EM}}/2  - \delta_{\mathrm {SU(2)}}/2 \right)~.
\end{equation}
Putting together the results on isospin breaking obtained in the 
previous section and the experimental values for the 
leptonic widths of the pion
$\Gamma_{\pi_{\mu 2 (\gamma)}} = 38.408 (7) \ (\mu s)^{-1}$~\cite{Nakamura:2010zzi} and the 
kaon
$\Gamma_{K_{\mu 2 (\gamma)}} = 51.25 (16) \ (\mu s)^{-1}$~\cite{Antonelli:2009ws} we obtain: 
\begin{eqnarray}
\frac{|V_{us}| F_K}{|V_{ud}| F_\pi}  & = &  0.23922(25)  \times 
\Bigg( \frac{\Gamma_{K_{\ell 2 (\gamma)}}}
{\Gamma_{\pi_{\ell 2 (\gamma)}}}
\Bigg)^{1/2} \nonumber \\
&=&   0.2763(5)~.
\end{eqnarray}
Finally, taking as reference value for $F_K/F_\pi$  the FLAG 
\cite{Colangelo:2010et} average of $N_f = 2 + 1$ calculations  
(see Eq. (\ref{eq:flag})) 
we find for the ratio of the CKM elements: 
\beq
\frac{|V_{us}|}{|V_{ud}|} =  0.2316(12)   ~.
\label{eq:VusKl2}
\eeq


As discussed in Ref. \cite{Antonelli:2010yf}, 
one can perform a fit to $|V_{ud}|$ and $|V_{us}|$   using as input  the 
values of  
$|V_{ud}| =   0.97425(22)$  from  superallowed nuclear $\beta$  
transitions~\cite{Hardy:2008gy}, 
$|V_{us}/V_{ud}|$  from $K_{\ell 2} / \pi_{\ell 2}$  
(Eq. (\ref{eq:VusKl2})),  
and $|V_{us}|$ from  $K_{\ell 3}$ decays.
Measurements of $K_{\ell 3}$ decays (BRs, form factors)~\cite{Antonelli:2010yf}, supplemented 
with theoretical calculations of electromagnetic and isospin corrections~\cite{Cirigliano:2008wn,Kastner:2008ch}
allow  one to extract  the product $|V_{us}| f_{+} (0) = 0.2163(5)$
\cite{Antonelli:2010yf}.
Combining this  with the $N_f = 2+1$  lattice result for the $K \to \pi$ vector form factor  
$f_+(0) = 0.9599 (34) (^{+31}_{-47}) (14)$~\cite{Boyle:2010bh} 
one obtains~\cite{Colangelo:2010et}:
\begin{equation}
|V_{us}| (K_{\ell 3}) = 0.2255(5)_{\rm exp}(12)_{\rm th}~.
\label{eq:VusKl3}
\end{equation}
A fit to the above described input leads to
\begin{equation}
|V_{ud} | =  0.97425(22)~, \quad
|V_{us}| =  0.2256(9)~,
\end{equation}
with $\chi^2/{\rm ndf} = 0.012 $ and negligible correlations between 
$|V_{ud}|$ and $|V_{us}|$. 
Fig.~\ref{fig:unitarity} provides a graphical representation of the various  
constraints in the  $|V_{us}| - |V_{ud}|$ plane and the $1 \sigma$ fit 
region (in yellow).  

The fit  values of $|V_{us}|$ and $|V_{ud}|$ (together with the negligible 
contribution from  $|V_{ub}| = 0.00393(36)$ \cite{Antonelli:2009ws})
imply a  stringent test of CKM unitarity: 
\beq
\Delta_{\rm CKM} = |V_{ud}|^2  +  |V_{us}|^2  +  |V_{ub}|^2  - 1  = 
  0.0001(6) ~.
\eeq
The remarkable agreement with the standard model prediction  $\Delta_{\rm 
CKM} = 0$ 
allows one to set strong  bounds on the effective scale of dimension-six operators  
that  violate quark-lepton universality of charged current weak interactions  \cite{Cirigliano:2009wk} 
and thus  parameterize new physics  contributions to $\Delta_{\rm CKM}$. 
The new physics effective  scale is constrained to be $\Lambda >  11$ TeV ($90 \%$ C.L.), 
a lower bound to be compared to the effective scale of the standard 
model set by the Fermi constant, 
$\Lambda_{\rm SM} = (2 \sqrt{2} G_F)^{-1/2} \simeq 174$ GeV. 
This  puts  this charged-current quark lepton universality 
test at the same level as the precision electroweak tests from 
$Z$-pole measurements.

\section{Conclusions}
\label{sec: Conclusions}
\renewcommand{\theequation}{\arabic{section}.\arabic{equation}}
\setcounter{equation}{0}

In this letter we have discussed in detail the effect of electromagnetic and strong isospin 
breaking  relevant for the extraction of $|V_{us}/V_{ud}|$  from the 
combination of the measured  $\Gamma_{K_{\mu 2 (\gamma)}} / \Gamma_{\pi_{\mu 2 (\gamma)}}$ 
and the ratio of decay constants $F_K/F_\pi$ calculated in lattice QCD 
{\it in the isospin limit}. 
We have performed our  analysis in the unified framework of ChPT, providing analytic 
expressions and numerical estimates for the EM and strong isospin breaking 
corrections 
$\delta_{\mathrm {EM}}$  and $\delta_{\mathrm {SU(2)}}$ that appear  in 
the 
ratio $\Gamma_{K_{\mu 2 (\gamma)}} / \Gamma_{\pi_{\mu 2 (\gamma)}}$ 
(see Eq. (\ref{modratio})). 

The main new result of our work is a compact analytic expression for the 
correction due to strong isospin breaking (see Eq. (\ref{deltaSU2v2})). 
This effect, neglected  in all previous extractions of $|V_{us}/V_{ud}|$,  
is comparable in size  to the electromagnetic correction:
\begin{equation}
\delta_{\mathrm {EM}} = -0.0069(17)~, \quad
\delta_{\mathrm {SU(2)}} = -0.0043(12)~. 
\end{equation}
Using the current experimental results for the decay widths 
and 
$F_K/F_\pi = 1.193(6)$~\cite{Colangelo:2010et}, 
we obtain $|V_{us}/V_{ud}| = 0.2316(12)$, 
which is about  $0.2 \%$ above previous determinations that 
ignored  $\delta_{\mathrm {SU(2)}}$.

\section*{Acknowledgements} 
We thank 
Gerhard Ecker, 
J\"urg Gasser, 
Gino Isidori, 
Andreas J\"uttner, and 
Heiri Leutwyler
for reading the manuscript and 
providing useful feedback.


\end{document}